\documentclass[12pt]{article}

\begin{document}

\title{Addendum to paper ``Closed Spaces in Cosmology''\cite{CSC}}
\author{Helio V. Fagundes\thanks{%
Instituto de F\'{i}sica Te\'{o}rica, Universidade Estadual Paulista, Rua
Pamplona, 145, S\~{a}o Paulo, SP 01405-900, Brazil. E-mail:
helio@ift.unesp.br}}
\maketitle

\begin{abstract}
A few corrections and comments are made upon a previously published paper,
on the subject of cosmological models with compact spatial sections.

\end{abstract}

The paper referred to in the title was published six years ago. Because of
growing interest in this field - see, for example, the review paper by
Lachi\`{e}ze-Rey and Luminet\cite{LaLu}, or, for a recent work, Levin et al.
\cite{LSS} - we find it relevant now to publish the present Adendum-Errata.

\bigskip

1) On page 203 of \cite{CSC}, a term is missing in equation (4). The correct
expression is

\qquad

\begin{center}
\[
Du^{a}=(\partial u^{a}/\partial x^{c}+\Gamma _{bc}^{a}u^{b})dx^{c}+(\partial
u^{a}/\partial z)dz 
\]
\end{center}

\bigskip

2) On p. 204, Table II, row 5: for type BVI($A),$ form $\omega ^{3}$ should
be $e^{(A+1)x}dz$.

\bigskip

3) Regarding Theorem 2.1, p. 204: \ \ ``If a closed space $M$ admits a BKS
metric, then $M$ is locally homogeneous with respect to this metric.'' In
the development of the proof in \cite{CSC}, it was stated that $M=\tilde{M}%
/\Gamma $, where $\Gamma \subset G$, with $G$ the corresponding BKS group.
As proved by Koike et al. \cite{JMP94}, this is not possible for type BVIII 
and all Bianchi types of class B. But the theorem remains valid, with $\Gamma $
a subgroup of the full group of the metric, Isom($\tilde{M})$.

Koike et al.'s result and Theorem 2.1 imply that a class B space can only
be compactified if its full (orientation preserving) isometry group has a
dimension larger than three. This explains why there are no closed spaces of
types BIV and BVI$(A\neq 0,1)$: their full isometry groups are of dimension
three, and so essentially coincide with their Bianchi groups.

\bigskip

4) On Table III, p. 206, make the following corrections:

\qquad \qquad a) for type BII, \qquad $\qquad K_{2}=K_{3}=+1/4$

\qquad\ \ \ \ \ \ b) for type BIV,\qquad $\qquad K_{1}=-3/4$

\qquad\ \ \ \ \ \ c) for type BVI$(A),\qquad \ d\lambda
^{2}=dx^{2}+e^{2(A-1)x}dy^{2}+e^{2(A+1)x}dz^{2}$

\bigskip

5) On p. 216, line 8, end of proof of Theorem 3.3: where is ``and hence in $%
\Sigma $,'' it should be ``and hence in $\tilde{\Sigma}$.''

\bigskip

6) On p. 216, paragraph beginning with ``Elementary particle theorists ...'':
the sentence \ \ ``This structure is not unique, but ... dimensions
involved.'' should be replaced with ``This structure is not unique, but we
can chooose one of the two alternatives as part of the \textit{definition}
of spinors. Cf. \cite{P&R}." 

\bigskip

I am grateful to Sandro Costa for calling my attention to the curvature
mistakes in Table III, and to Conselho Nacional de Desenvolvimento
Cient\'{i}fico e Tecnol\'{o}gico (CNPq - Brazil) for partial financial
support.

\bigskip

\end{document}